\documentstyle[12pt,graphicx]{article}
\title{On Kelvin-Helmholtz instability in superfluids}
\author{G.E. Volovik
\\Low Temperature Laboratory, Helsinki University of
Technology\\
P.O.Box 2200, FIN-02015 HUT, Finland\\
\\
 L.D. Landau Institute for
Theoretical Physics\\  Kosygin Str. 2, 117940 Moscow, Russia
}

\begin{document}
\maketitle

\abstract{The  Kelvin-Helmholtz instability  in superfluids is
discussed,
based on the first experimental observation of such instability at the
interface between superfluid $^3$He-A and superfluid $^3$He-B
(R.~Blaauwgeers, V.\,B.~Eltsov, G.~Eska et al.,
cond-mat/0111343)
[7]. We discuss  why the
Kelvin-Helmholtz criterion, the Landau critical velocity for
nucleation of
ripplons, and the free energy consideration all give different values
for the
instability treshold.}
\vskip 5mm

PACS: 47.20.Ma, 67.57.Np, 68.05.$-$n

\vskip 5mm

\maketitle

{\bf 1.~Classical Kelvin-Helmholtz (KH) instability.}
KH instability
belongs to a broad class of interfacial instabilities in liquids,
gases,
plasma, etc.
\cite{Birkhoff}. It refers to the dynamic instability of the
interface of
the discontinuous flow, and may be defined  as the instability of the
vortex
sheet. Many natural phenomena have been attributed to this
instability. Most
familiar of them  are generation by wind of waves in the water, whose
Helmholtz instability \cite{Helmholtz} was first analyzed by Kelvin
\cite{LordKelvin}, and flapping of sails and flags analyzed by
Rayleigh
\cite{Rayleigh} (see recent experiments in \cite{FlexibleFilament}).

Many of the leading ideas in the theory of instability were originally
inspired by considerations about inviscid flows.  The corrugation
instability of the interface between two ideal liquids sliding with
along
each other was first investigated by Lord Kelvin
\cite{LordKelvin,ThomsonLandauLifshitz}.
The critical relative velocity  $|v_1-v_2|$
for the onset of corrugation instability is given by
\begin{equation}
 \frac{1}{ 2}  \frac{\rho_1\rho_2}{ \rho_1+\rho_2} (v_1-
v_2)^2=\sqrt{\sigma
 F}~.
 \label{InstabilityCondition1}
 \end{equation}
Here $\sigma$ is surface
tension of the interface between two liquids; $\rho_1$ and $\rho_2$
are their
mass densities; and  $F$ is related to the external field stabilizing
the
position of the interface: typically it is the gravitational field
\begin{equation}
F=g(\rho_1-\rho_2)~.
\label{GravityForce}
\end{equation}
The surface mode (ripplon) which is excited first has the wave vector
\begin{equation}
k_0=\sqrt{F/\sigma}~,
\label{WaveVectorInstability}
\end{equation}
and frequency
\begin{equation}
 \omega_0=k_0 \frac{\rho_1 v_1+\rho_2 v_2}{ \rho_1+\rho_2} ~.
\label{FrequencyInstability}
\end{equation}
The excited  ripplon
propagates along the interface with the phase and group velocity:
$v_{\rm
ripplon}=(\rho_1 v_1+\rho_2 v_2) / (\rho_1+\rho_2)$.

However, among the ordinary
liquids one cannot find an ideal one. That is why in ordinary liquids
and
gases it is not easy to correlate theory with experiment. In
particular,
this is because one cannot properly prepare the initial state -- the
planar
vortex sheet is never in equilibrium in a viscous fluid: it is not the
solution of the hydrodynamic equations if viscosity is finite. That
is why
it is
not so apparent whether one can properly discuss its `instability'.

Superfluids are the only proper ideal objects  where these ideas can
be
implemented without reservations, and where the criterion of
instability
does not contain viscosity. Recently the first experiment has been
performed in superfluids, where the nondissipative initial state was
well
determined, and the well defined treshold has been reported
\cite{Kelvin-HelmholtzInstabilitySuperfluids}. The initial state is
the
nondissipative vortex sheet separating two sliding superfluids. One
of the
superfluids ($^3$He-A) performs the solid-body like rotation together
with the
vessel, while in the other one  ($^3$He-B) the superfluid component
is in the
so-called Landau state, i.e. it is vortex-free and thus is stationary
in the
inertial frame. The threshold of the Kelvin-Helmholtz type
instability has been
marked by formation of vortices in the vortex-free stationary
superfluid: this
initially stationary superfluid starts to spin-up  by the neighboring
rotating
superfluid.

\smallskip
{\bf2.~KH instability in superfluids at low $T$.}
The extension of the consideration of classical KH instability to
superfluids adds some new physics. First of all, it is now the two-
fluid
hydrodynamics with superfluid and normal components which must be
incorporated. Let us first consider the limit case of low $T$, where
the
fraction of the normal component is negligibly small, and thus the
complication of the two-fluid hydrodynamics is avoided. In this case
one may
guess that the classical result (\ref{InstabilityCondition1})
obtained for
the ideal inviscid liquids is applicable for superfluids too, and the
only
difference is that the role of the gravity is played by the applied
gradient
of magnetic field $H$, which stabilizes the position of the interface
between
$^3$He-A and
$^3$He-B in the experiment \cite{Kelvin-
HelmholtzInstabilitySuperfluids}:
\begin{equation}
F= (1/2)(\chi_A(T) -\chi_B(T))\nabla (H^2) ~.
\label{InstabilityCondition2}
\end{equation}
Here $\chi_A$ and $\chi_B$ are temperature dependent magnetic
susceptibilities of the A and B phases.

However, this is not the whole story. The instability will start
earlier,
if one takes into account that there is a preferred reference frame.
It can
be  the frame of container, the frame of the crystal in
superconductors, or
even the frame where the inhomogeneity of magnetic field $H$ is
stationary.
The  energy of the excitations of the surface, ripplons, can become
negative
in this reference frame, and the surface modes will be excited,
before the
onset of the classical KH instability.

Let us consider this phenomenon.
We repeat the same derivation as in case of classical KH instability,
assuming the same boundary conditions, but with one important
modification:
in the process of the dynamics of the interface one must add the
friction
force arising when the interface is moving with respect to the
container
wall. In the frame of the container, which
coincides with the frame of the stable position of the interface, the
friction force between the interface and container is
\begin{equation}
F_{friction}=-\Gamma \partial_t\zeta~,
\label{FrictionForceInterface}
\end{equation}
 where $\zeta(x,t)$ is perturbation of the position of the interface:
\begin{equation}
 z=z_0+\zeta(x,t)~,~\zeta(x,t)=a\sin (kx-\omega t)~.
\label{zeta}
\end{equation}
We assume that the velocities $v_1$ and
$v_2$ are both along the axis
$x$; the container walls are parallel to the $(x,z)$-plane; and the
interface is
parallel to the $(x,y)$-plane.

The friction force in
Eq.(\ref{FrictionForceInterface}) violates the Galilean invariance in
$x$-direction, which reflects the existence of the preferred
reference frame
-- the frame of container. This symmetry breaking is the main reason
of the
essential modification of the KH instability.  The parameter $\Gamma$
in the
friction force has been calculated for the case when the interaction
between
the interface and container is transferred by the normal component of
the
liquid due to Andreev scattering of
ballistic quasiparticles by the interface \cite{KopninInterface}.
The friction modifies the classical spectrum of surface modes:
\begin{equation}
  \rho_1\left(\frac{\omega}{
k}-v_1\right)^2+\rho_2\left(\frac{\omega}{
k}-v_2\right)^2
=
\frac{F+k^2\sigma}{k} -i\Gamma\frac{\omega}{ k}~,
\label{SpectrumNonzeroFrictionZeroT}
\end{equation}
or
\begin{eqnarray}
&\displaystyle
\frac{\omega}{
k}=  \frac{\rho_1 v_1+\rho_2 v_2}{\rho_1+\rho_2}\pm &\nonumber\\
&\displaystyle\pm
\frac{1}{
\sqrt{\rho_1+\rho_2}}
\sqrt{
\frac{F+k^2\sigma}{k} -i\Gamma\frac{\omega}{k} - \frac{\rho_1 \rho_2
}{
\rho_1+\rho_2} (v_1-v_2)^2} ,&
\label{SpectrumNonzeroFrictionZeroT2}
\end{eqnarray}
where $v_1$ and $v_2$ are the velocities of superfluid components of
the
liquids with respect to the container walls.

For
$\Gamma=0$ the spectrum of ripplons acquires the imaginary part,
${\rm Im}\omega(k)\neq 0$, at the classical treshold value in
Eqs.(\ref{InstabilityCondition1}) and (\ref{WaveVectorInstability}).
However, the frame-fixing parameter $\Gamma$ leads to essentially
different
result: The imaginary part of frequency becomes positive
${\rm Im}\omega(k)>0$ first for ripplons with the
same value of the wave vector, as in
Eq.(\ref{WaveVectorInstability}), but
the ripplon frequency is now $\omega=0$ and its group velocity is
$v_{group}=d\omega/dk=0$. The critical ripplon is stationary in the
reference
frame of the container, as a result the onset of instability is given
 by
\begin{equation}
 \frac{1}{2}   \rho_1 v_1^2 + \frac{1}{2} \rho_2v_2^2=\sqrt{\sigma
F}~.
\label{InstabilityConditionNew}
\end{equation}
This criterion does not depend on relative velocities of superfluids,
but is
determined by velocities of each of the two superfluids with respect
to
the container (or to the remnant normal component).  Thus the
instability can
occur even if two liquids have equal densities, $\rho_1=\rho_2$, and
move
with the same velocity, $v_1=v_2$.  This situation is very similar to
the
phenomenon of flapping flag in wind, discussed by Rayleigh in terms
of the KH
instability -- the instability of the passive deformable membrane
between two
distinct parallel streams having the same density and the same
velocity (see
latest experiments in Ref.  \cite{FlexibleFilament}). In our case the
role of
the flag is played by the interface, while the role of the flagpole
which
pins the flag (and thus breaks the Galilean invariance) is played by
the
container wall.

Note that in the limit of the vanishing pinning parameter
$\Gamma\rightarrow
0$ the Eq.(\ref{InstabilityConditionNew})  does
not coincide with the classical equation (\ref{InstabilityCondition1})
obtained when there is no pinning, i.e. when $\Gamma$ is exactly
zero. Such
difference between the limit and exact cases is known in many area of
physics. In classical hydrodynamics the normal mode of inviscid
theory may
not be the limit of a normal mode of viscous theory \cite{LinBenney}.
Below
we discuss this difference for the case of KH instability in
superfluids.

\smallskip
{\bf 3.~KH instability and modified Landau criterion.}
Let us first compare both results, with no pinning ($\Gamma=0$) and
for
vanishing
pinning
($\Gamma\rightarrow 0$),  with the Landau criterion. According to
Landau, a
quasiparticle is created by the moving superfluid if its velocity
 with respect to the container wall (or with respect to the
body moving in superfluid) exceeds
\begin{equation} v_{\rm Landau}=\min
\frac{E(p)}{p}  ~.
\label{Landaucriterion}
\end{equation}
Let us recall that
  the energy $E(p)$ here is the   quasiparticle  energy in the
reference
frame  moving with the superfluid vacuum. In our case there are two
superfluids moving with different velocities.  That is why there is
no unique
superfluid comoving frame, where $E(p)$ can be uniquely determined.
Such
frame appears only in particular cases, when either $v_1=v_2$, or if
instead
of the interface one considers the free surface of a single liquid
(i.e. if
$\rho_2=0$). In these particular cases the Landau criterion in the
form of
Eq.(\ref{Landaucriterion})  must work.  The energy spectrum of the
ripplons
on the interface between two stationary fluids (or on the surface of
the
single liquid) is given by Eq.(\ref{SpectrumNonzeroFrictionZeroT2})
with
$v_1=v_2=\Gamma=0$:  \begin{equation} \frac{\omega^2(k)}{k^2}=
\frac{1}{
\rho_1+\rho_2}
\frac{F+k^2\sigma}{k}   ~.
\label{RipplonSpectrum}
\end{equation}
This gives the following Landau  critical velocity:
\begin{equation}
v^2_{\rm Landau}=\min \frac{\omega^2(k)}{
k^2}= \frac{2}{
\rho_1+\rho_2}\sqrt{ F \sigma}~.
\label{LandauVelocityInterface}
\end{equation}
This coincides with the Eq.({\ref{InstabilityConditionNew}) if
$v_1=v_2$, or
if $\rho_2=0$. But this does not coincide with the classical KH
result: the
latter is obtained at $\Gamma=0$ when the interaction with the
reference
frame of
the container is lost, and thus the Landau criterion is not
applicable.

In the general case, when neither of the two conditions ($v_1=v_2$, or
$\rho_2=0$)
fulfils, the Landau criterion must be reformulated:
the instability occurs, when  the frequency of the surface mode
in the frame of the container crosses zero for the first time:
$\omega(k;v_1,v_2)=0$.  Inspection of
Eq.(\ref{SpectrumNonzeroFrictionZeroT2}) with $\Gamma=0$ shows that
for
$k=k_0$ the spectrum with negative square root touches zero just
when the treshold (\ref{InstabilityConditionNew}) is reached. Thus the
Landau criterion in its general formulation coincides with the
criterion
of instability obtained for the case of  nonzero friction force.  As
distinct
from the Landau criterion in the form of (\ref{Landaucriterion})
valid for
a single superfluid velocity, where it is enough to know the ripplon
spectrum
in the frame where the superfluid\,($s$) is (are) at rest, in the
general case
one must calculate the ripplon spectrum $\omega(k;v_1,v_2)$ for the
relatively moving superfluids.

\smallskip
{\bf 4.~Matching zero-pinning and vanishing-pinning regimes.}
The difference in the result for onset of KH instability in the two
regimes
-- with $\Gamma=0$ and with $\Gamma\neq 0$ -- disappears only in the
case
when two
superfluids move in such a way that in the reference
frame of container the combination $\rho_1v_1+\rho_2v_2 =0$. In this
arrangement, according to Eq.(\ref{FrequencyInstability}), the
frequency of
the ripplon created by classical KH instability is zero in the
container
frame. Thus at this special condition the two criteria,
zero pinning (\ref{InstabilityCondition1}) and
vanishing pinning (\ref{InstabilityConditionNew}), must coincide; and
they
really
do.

If $\rho_1v_1+\rho_2v_2 \neq
0$, the crossover between the zero pinning regime and the regime of
small
pinning
occurs by varying the observation time.  Let us consider this on the
example of
the experimental set-up
\cite{Kelvin-HelmholtzInstabilitySuperfluids} with the vortex-free B-
phase
and the vortex-full A-phase in the
rotating vessel: In the
container frame one has
${\bf v}_1= {\bf v}_{{s}A}=0$~,~
${\bf v}_2= {\bf v}_{{s}B}=-{\bf\Omega}\times{\bf r}$; the densities
of
two liquids,
$^3$He-A and $^3$He-B, are the same with high accuracy:
$\rho_A=\rho_B=\rho$.
In the non-zero pinning regime the instabilty occurs at the boundary
of the
vessel,
where the velocity of the $^3$He-B is maximal, when this maximal
velocity
reaches
the value:
\begin{equation}
 v_c^2=\frac{2}{\rho}\sqrt{F\sigma}=\frac{1}{2}v_{\rm KH}^2=2v_{\rm
 Landau}^2~.  \label{CriticalVelocityRotatingVessel}
 \end{equation}
 This
velocity is by $\sqrt{2}$ smaller than that given by classical KH
equation
(\ref{InstabilityCondition1}) for the zero-pinning regime. On the
other hand
it is by $\sqrt{2}$ larger than the Landau criterion in the form of
Eq.(\ref{Landaucriterion}), but coincides with Landau criterion
properly
formulated for two superfluids.

From Eq.({\ref{SpectrumNonzeroFrictionZeroT}) it follows that
slightly above this treshold the increment of the exponential  growth
of the
interface perturbation is
\begin{equation}
{\rm Im}~\omega(k_0)=   \frac{ \Gamma k_0}{ 2\rho}\left(\frac{v_{{
s}B}}{
v_c}-1\right)~~,~{\rm at}~ ~ v_{{s}B} -v_c  \ll v_c ~.
\label{Increment}
\end{equation}
In the vanishing pinning limit $\Gamma\rightarrow 0$  the increment
becomes small and the discussed instability of the surface has no
time for
development if the observation time is short enough. It will start
only at
higher
velocity of rotation when the classical treshold of KH instability,
$v_{KH}$ in Eq.(\ref{InstabilityCondition1}), is reached. Thus,
experimental
results in this limit would depend on the observation time -- the time
one waits for the interface to be coupled to the laboratory frame and
for the
instability to develop. For sufficiently short time one will measure
the
classical
KH criterion (\ref{InstabilityCondition1}), while for the
sufficiently long
observation time the modified KH criterion
(\ref{CriticalVelocityRotatingVessel})
will be observed.

\smallskip
{\bf 5.~Thermodynamic instability.}
\label{ThermodynamicInstabilitySec}
Let us now consider the case of nonzero $T$, when each of the two
liquids
contain superfluid and normal components. In this case the analysis
requires
the $2\times 2$-fluid hydrodynamics. This appears to be rather
complicated
problem, taking into account that in some cases the additional
degrees of
freedom related to the interface itself must be also added. The two-
fluid
hydrodynamics has been used for investigation of the instability of
the free
surface of superfluid $^4$He by the relative motion of the normal
component
of the
liquid with respect to the superfluid one \cite{Korshunov}.  We avoid
all these
complications assuming that the viscosity of the normal components of
both
liquids
is high, as it actually happens in superfluid
$^3$He. In this high-viscosity limit we can neglect the dynamics of
the normal
components, which is thus fixed by the container walls. Then the
problem is
reduced
to the problem of the thermodynamic instability of the superflow in
the
presence
of the interface.

We start with the
following initial non-dissipative state corresponding to the thermal
equilibrium in the presence of the interface and superflows. In
thermal
equilibrium the normal component must be at   rest in the container
frame,
${\bf v}_{{ n}1}={\bf v}_{{ n}2}=0$, while the superfluids can move
along the interface with velocities
${\bf v}_{{ s}1}$ and ${\bf v}_{{ s}2}$ (here the velocities are in
the
frame of the container).

The onset of instability can be found from
free-energy consideration: When the free energy of
static perturbations of the interface becomes negative in the frame
of the
container, the initial state becomes thermodynamically unstable. The
free-energy
functional for the  perturbations of the interface in the reference
frame
of the
container  is determined by `gravity', surface tension, and
perturbations
$\tilde{\bf v}_{{ s}1}=\nabla\Phi_{1}$ and $\tilde{\bf v}_{{
s}2}=\nabla\Phi_{2}$ of the velocity field caused by deformation of
the
interface:
\begin{eqnarray}
&\displaystyle
{\cal F}\{\zeta\}=\frac{1}{2}\int dx \Biggl(F\zeta^2 + \sigma
(\partial_x\zeta)^2 +{}&\nonumber\\
&\displaystyle
{}+
\int_{-\infty}^\zeta dz \rho_{{ s}1ik} \tilde v_{{ s}1}^i
\tilde v_{{ s}1}^k +
\int_\zeta^{\infty} dz \rho_{{ s}2ik} \tilde v_{{ s}2}^i \tilde v_{{
s}2}^k\Biggr).&
\label{SurfaceFunctionalAnisotropic}
\end{eqnarray}
For generality we discuss anisotropic
superfluids, whose superfluid densities are tensors (this occurs in
$^3$He-A). The
velocity perturbation fields
$\tilde v_{{ s}k}=\nabla\Phi_k$, obeying the continuity equations
$\partial_i(\rho_{ s}^{ik} \tilde v_{{ s}k})=0$,  have the following
form:
\begin{eqnarray}
&\begin{array}{c}
 \Phi_1(x,z<0)=A_1e^{k_1z}\cos  kx ,\\
\Phi_2(x,z>0)=A_2e^{-k_2z}\cos  kx, \end{array}&
\label{PhiAnisotropic}\\
&\rho_{{ s}1z}
k_1^2 =\rho_{{ s}1x} k^2~,~\rho_{{ s}2z} k_2^2 =\rho_{{ s}2x} k^2.&
\label{kAnisotropic}
\end{eqnarray}
The connection between the deformation of the surface,
$\zeta(x)=a\sin  kx$,
and the velocity perturbations follow from the boundary conditions.

Because of large viscosity of the normal component it is clamped by
the
boundaries of the vessel. Then from the requirement that the mass and
the
heat currents are conserved across the wall, one obtains that the
superfluid
velocity in the direction normal to the wall must be zero: ${\bf
v}_{{ s}1}\cdot {\bf n}={\bf v}_{{ s}2}\cdot {\bf n}=0$. This gives
the following boundary conditions for perturbations:
\begin{equation}
\partial_z\Phi_1=   v_{{ s}1} \partial_x\zeta ,~~~\partial_z\Phi_2=
v_{{ s}2}
\partial_x\zeta  ~.
\label{NoPenetration1}
\end{equation}
Substituting this to the free-energy functional
(\ref{SurfaceFunctionalAnisotropic}), one obtains  the   quadratic
form
of the free energy of the surface modes
\begin{eqnarray}
&\displaystyle
{\cal F}\{\zeta\}= \frac{1}{2}\sum_k
|\zeta_k|^2\times{}&\nonumber\\
&\displaystyle
\times
\left(F + k^2\sigma  -k
\left(\sqrt{\rho_{{ s}x1}\rho_{{
s}z1}} v_{{ s}1}^2+ \sqrt{\rho_{{ s}x2}\rho_{{
s}z2}} v_{{ s}2}^2  \right)\right)&
\label{SurfaceFunctionalAnisotropic2}
\end{eqnarray}
This energy becomes negative for the first time for the mode with
$k_{0}=(F/\sigma)^{1/2}$ when
\begin{equation}
 \frac{1}{2}  \left(\sqrt{\rho_{{ s}x1}\rho_{{
s}z1}} v_{{ s}1}^2+ \sqrt{\rho_{{ s}x2}\rho_{{
s}z2}} v_{{ s}2}^2  \right)=\sqrt{\sigma
F}~.
\label{InstabilityConditionNewNonzeroT}
\end{equation}
This is the criterion (\ref{InstabilityConditionNew}) for the non-
zero pinning
regime extended to finite temperatures.
Eq.(\ref{InstabilityConditionNewNonzeroT}) transforms to
Eq.(\ref{InstabilityConditionNew}) when $T\rightarrow 0$: The normal
components of the liquids disappear and one has $\rho_{{ s}x1}=
\rho_{{ s}z1}=\rho_1$ and $\rho_{{ s}x2}=
\rho_{{ s}z2}=\rho_2$.

\smallskip
{\bf 6.~Nonlinear stage of instability.}
Eq.(\ref{InstabilityConditionNewNonzeroT}) is in excelent agreement
with the
onset of the surface instability measured in experiment
\cite{Kelvin-HelmholtzInstabilitySuperfluids}. The onset of
instability is marked by the appearance of the vortex lines in $^3$He-
B which
are monitored in NMR measurements.  This demonstrates that
vortices appear in the nonlinear stage of this KH instability.

The precise mechanism of the vortex formation is not yet known. One
may guess
that the A-phase vorticity is pushed by the Magnus force towards the
vortex-free B-phase region \cite{MovingAB}. When the potential well
for
vortices is formed by the corrugation of the interface (see Figure),

\begin{figure}
  \centerline{\includegraphics[width=1.0\linewidth]{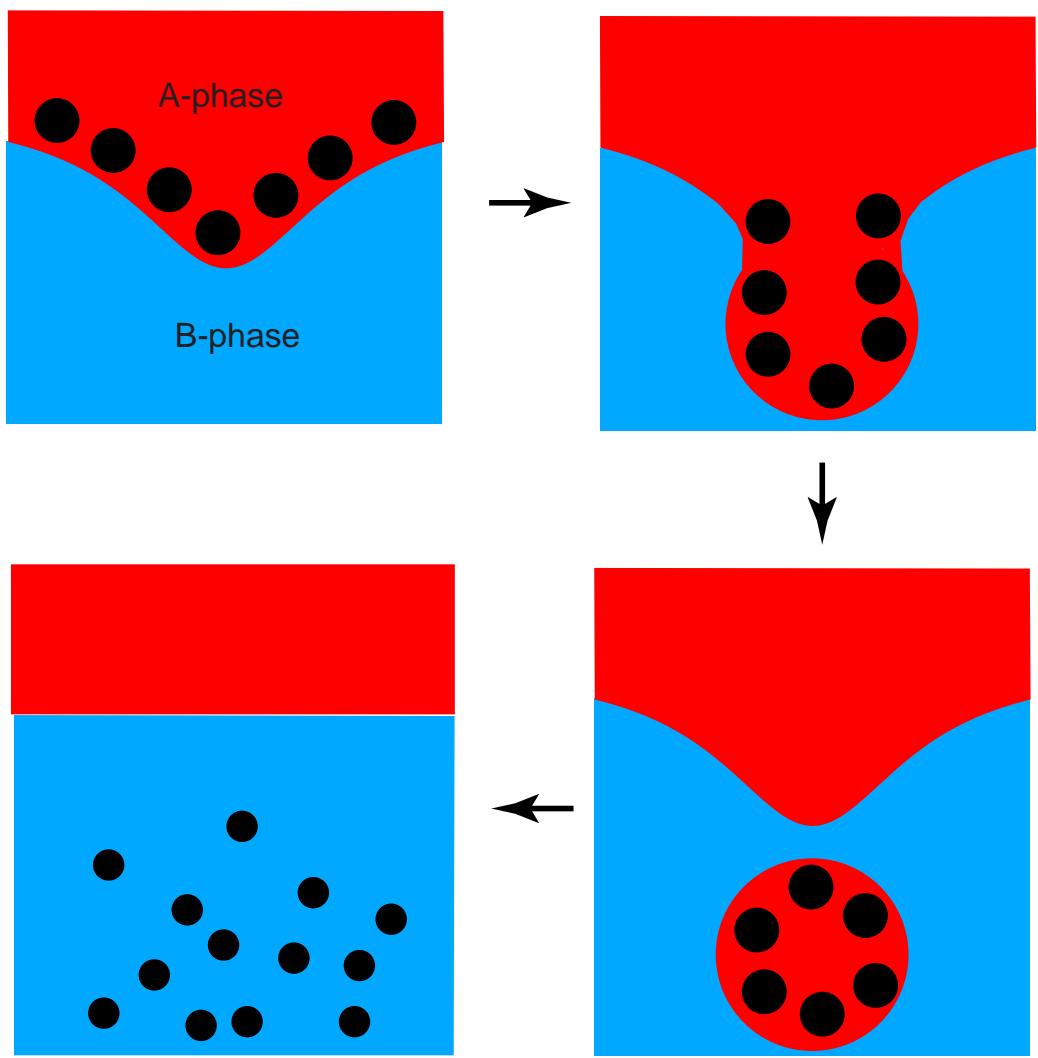}}
  \caption{Possible scenario of vortex formation by Kelvin-Helmholtz
instability of
the AB interface. }
  \label{KelvinInstabilityFig}
\end{figure}

the vortices are pushed there and enhance
further the growth of the  potential well, until it forms the
 droplet of the A-phase filled by vorticity. The vortex-full droplet
propagates to
the bulk B-phase where it relaxes to the singular vortex
lines of
$^3$He-B.

Under the conditions of the experiment the nucleation of vortices
leads
to decrease of the B-phase velocity below the instability treshold,
and the
vortex
formation is stopped. That is why one may expect that  the vortex-full
droplet is
nucleated during the development of the instability from a single
seed. The
size of
the seed is about one-half of the  wavelength
$\lambda_0=2\pi/k_0$ of the perturbation. The number of the created
vortices  is
found from the circulation of superfluid velocity carried by the
piece  of the
vortex sheet of size $\lambda_0/2$, which is determined by the jump of
superfluid
velocity across the sheet:
$\kappa=|{\bf v}_{{
 s}B}-{\bf v}_{{  s}A}|\lambda_0/2$. Dividing this by the circulation
quantum
$\kappa_0$ of the created B-phase vortices one obtains the number of
vortices
produced as the result of the growth of one segment of the
perturbation:
\begin{equation}
N=\frac{\kappa}{\kappa_0} \sim  \frac{v_c\lambda_0 }{2\kappa_0}~.
\label{NumberCreatedVortices}
\end{equation}
It is about 10 vortices per event under condition of the experiment,
which is
in a good agreement with the measured number of vortices created per
event
\cite{Kelvin-HelmholtzInstabilitySuperfluids}. This is in favour of
the droplet
mechanism of vortex formation.

Probably, the experiments on KH instability in superluids will allow
to
solve the
similar problem of the non-linear stage of instability in ordinary
liquids
(see, for example, Ref. \cite{Kuznetsov}).

The vortex formation by surface instability is rather generic
phenomenon. This
mechanism has been discussed for vortex formation in the laser
manipulated Bose
gases
\cite{VorBEC,DynamicInstabilityBEC}.  It can be applicable to
different
kinds of
interfaces, and  under very different
physical conditions. In particular, vortices can be generated at the
second order phase boundary  between the   normal and the
superfluid phases
\cite{Aranson}. Such an interface naturally appears at the rapid phase
transition into the superfluid state \cite{BigBangSimulation}.
The instability of the free surface of superfluid under the relative
flow of
the normal and superfluid components of the same liquid has been
recently
reexamined by Korshunov \cite{KorshunovNew}. He also obtained two
criteria
of instability: for zero and nonzero values of the viscosity of the
normal
component of the liquid.

I thank R.\,Blaauwgeers, V.\,B.\,Eltsov,  N.\,Ino\-ga\-mov,
N.\,B.\,Kopnin,
S.\,E.\,Korshunov, M.\,Krusius, E.\,A.\,Kuz\-net\-sov,  and E.V.
Thuneberg for
fruitful discussions. This work was supported by ESF COSLAB Programme
and by
the Russian Foundations for Fundamental Research.


\begin{thebibliography}{15}


\bibitem{Birkhoff} G. Birkhoff, Helmholtz and Taylor instability, in:
{\it Hydrodynamic Instability}, Proc.of Symposia in Appl. Mathematics, Vol.
XIII,  Edited by G. Birkhoff, R. Bellman and C.C. Lin, American Mathematical
Society 1962, pp.55-76.

\bibitem{Helmholtz} H.L.F. von Helmholtz, \"Uber discontinuierliche
Fl\"ussigkeitsbewegungen, Monatsberichte der k\"onigl. Akad. der
Wissenschaften zu Berlin, 215-228 (1868).

\bibitem{Rayleigh} Lord Rayleigh (J.W. Strutt), {\it Scientific papers},
Vol. {\bf 1}, Cambridge University Press, 1899.

\bibitem{FlexibleFilament} J. Zhang, S. Childress, A. Ubchaber, M. Shelley,
Flexible filaments in a flowing soap film as a model of one-dimensional flags
in a two-dimensional wind, Nature {\bf  408},  835-839 (2000).

\bibitem{LordKelvin} Lord Kelvin (Sir William Thomson) {\it Mathematical
and physical papers}, Vol. {\bf 4}, {\it Hydrodynamics and General
Dynamics}, Cambridge University Press, 1910.


\bibitem{ThomsonLandauLifshitz} W. Thomson (1871) in: L.D. Landau, and
E.M. Lifshitz, {\it Fluid Mechanics}, Pergamon Press, 1989, Sec. 62 `Capillary
waves', problem 3, page 247.

\bibitem{Kelvin-HelmholtzInstabilitySuperfluids} R. Blaauwgeers, V.B.
Eltsov, G. Eska, A.P. Finne,  R.P.
Haley, M. Krusius, J.J. Ruohio, L. Skrbek, and G.E. Volovik, Vortex lines at a
phase boundary between different quantum vacua,
cond-mat/0111343 v.1; Shear flow and Kelvin-Helmholtz instability in
superfluids, cond-mat/0111343 v.2.


\bibitem{LinBenney} C.C. Lin, and D.J. Benney, On the instability of shear
flows, in: {\it Hydrodynamic Instability}, Proc.of Symposia in Appl.
Mathematics, Vol. XIII,  Edited by G. Birkhoff, R. Bellman and C.C. Lin,
American Mathematical Society 1962, pp.1-24.

\bibitem{KopninInterface} N.B. Kopnin,   Movement of the interface
between the A and B phases os superfluid helium-3: linear theory ,
ZhETF, {\bf 92}, 2106 (1987);
\lbrack   Sov. Phys. JETP {\bf 65}, 1187 (1987) \rbrack;
 A.J. Leggett and S. Yip, Nucleation and
growth of $^3$He-B on the supercooled A-phase, in: {\bf Helium Three}, eds.
W.P.Halperin, L.P.Pitaevskii, Elsevier Science Publishers B.V., p. 523 (1990);
S. Yip and A.J. Leggett,   Dynamics of the
$^3$He A-B interface,   Phys. Rev. Lett., {\bf 57}, 345 (1986); J. Palmeri,
Superfluid kinetic equation approach to the dynamics of the
$^3$He A-B phase boundary, Phys. Rev. {\bf B~42}, 4010­4035 (1990).

\bibitem{Korshunov} S.E. Korshunov, Instability of superfluid helium free
surface
in the presence of heat flow, Europhys. Lett. {\bf 16}, 673-675 (1991).

\bibitem{MovingAB}
M. Krusius,   E.V.  Thuneberg, and \"U. Parts, AB phase transition in rotating
superfluid $^3$He,   Physica B  {\bf 197,} 367--389 (1994);
 \"U. Parts, Y. Kondo, J.S. Korhonen, M. Krusius, and  E.V.  Thuneberg,
Vortex layer at the interface between the $A$ and $B$ phases of superfluid
$^3$He, Phys. Rev. Lett. {\bf 71}, 2951--2954 (1993).

\bibitem{Kuznetsov} E.A. Kuznetsov, and P.M. Lushnikov, Nonlinear theory of
excitation of waves by Kelvin-Helmholtz instability, ZhETF {\bf 108}, 614-630
(1995).

\bibitem{VorBEC}
K.W.  Madison, F. Chevy, V. Bretin, and J. Dalibard,
  Stationary states of a rotating Bose-Einstein condensate: Routes to vortex
  nucleation,   Phys. Rev. Lett.  {\bf 86,} 4443--4446 (2001).

\bibitem{DynamicInstabilityBEC}  S. Sinha and Y. Castin,  Dynamic
instability of a rotating Bose-Einstein condensate,
Phys. Rev. Lett. {\bf 87}, 190402 (2001).


\bibitem{Aranson}  I. S. Aranson,  N. B. Kopnin,  and V. M. Vinokur,
Dynamics of vortex nucleation by rapid thermal quench, Phys.
Rev. {\bf B~ 63}, 184501 (2001); Nucleation of vortices in superfluid
$^3$He-B by
rapid thermal quench, in:
{\it Vortices in unconventional superconductors and superfluids}, eds. R.P.
Huebener, N. Schopohl, and G.E. Volovik, Springer Series in Solid-State
Science {\bf 132}, Springer, 2002, pp. 49-64.

\bibitem{BigBangSimulation} V.M.H. Ruutu, V.B. Eltsov, A.J. Gill, T.W.B.
Kibble, M. Krusius, Yu.G. Makhlin, B. Placais, G.E. Volovik, and Wen Xu,
Vortex
formation in neutron-irradiated superfluid 3He as an analogue of cosmological
defect formation,   Nature, {\bf 382}, 334-336 (1996)

\bibitem{KorshunovNew}
S. E. Korshunov, Kelvin-Helmholtz instability of superfluid liquid free
surface,  cond-mat/0203374.


\end{thebibliography}
\end{document}